\documentclass[english,aps,prl,twocolumn,superscriptaddress,showpacs]{revtex4}
\usepackage{graphicx}
\usepackage{babel}

\begin{document}

\title{Impact of observational uncertainties on universal scaling of MHD turbulence}

\author{G.~Gogoberidze}
\email{g.gogoberidze@warwick.ac.uk}
\affiliation{Centre for Fusion, Space and Astrophysics; University of Warwick,
Coventry, CV4 7AL, United Kingdom}
\affiliation{Institute of Theoretical Physics, Ilia State University, 3/5 Cholokashvili ave., 0162 Tbilisi, Georgia}

\author{S.C.~Chapman}
% \email{s.c.chapman@warwick.ac.uk}
\affiliation{Centre for Fusion, Space and Astrophysics; University of Warwick,
Coventry, CV4 7AL, United Kingdom}

\author{B.~Hnat}
% \email{b.hnat@warwick.ac.uk}
\affiliation{Centre for Fusion, Space and Astrophysics; University of Warwick,
Coventry, CV4 7AL, United Kingdom}

\author{M.W.~Dunlop}
\affiliation{Centre for Fusion, Space and Astrophysics; University of Warwick,
Coventry, CV4 7AL, United Kingdom}
\affiliation{Rutherford-Appleton Laboratory, Chilton, Oxfordshire, OX11 0QX, United Kingdom}
\affiliation{CSSAR, CAS, PO Box 8701, Beijing 100190, China}
\affiliation{The Blackett Laboratory, Imperial College London, London, SW7 2AZ, United Kingdom}

\begin{abstract}
Scaling exponents are the central quantitative prediction of theories of turbulence and in-situ satellite observations of the high Reynolds number solar wind flow have provided an extensive testbed of these.
%These observations are challenging and their uncertainties are nontrivial to determine.
We propose a general, instrument independent method to estimate the uncertainty of velocity field fluctuations.
%This uncertainty is scale dependent and
We obtain the systematic shift that this uncertainty introduces into the observed spectral exponent. This shift is essential for the correct interpretation of observed scaling exponents. It is sufficient to explain the contradiction between spectral features of the Elsasser fields observed in the solar wind with both theoretical models and numerical simulations of Magnetohydrodynamic turbulence.
\end{abstract}

\pacs{94.05.Lk, 52.35.Ra, 95.30.Qd, 96.60.Vg}

\maketitle

Universality in isotropic, homogeneous turbulence is expressed through its statistical scaling properties. In the absence of intermittency, the scaling exponent for the  inertial interval of hydrodynamic turbulence is completely determined by the assumption of self-similarity
\cite{K41}, leading to the well known unique $-5/3$ power spectral exponent.
This is not the case for Magnetohydrodynamic (MHD) turbulence where the magnetic field introduces an additional  physical quantity with the dimension of velocity, namely, the Alfv\'en velocity \cite{CEa08} and indeed it is an open question as to whether the scaling is universal. Detailed phenomenological models of MHD turbulence are thus needed to predict the scaling exponent, and its precise observational determination is essential in order to validate these theories.

In-situ satellite observations of the solar wind magnetic field and bulk flow span several decades in temporal scales and offer a 'natural laboratory'  for the study of MHD turbulence, the Reynolds number  exceeds $\sim 10^5$ \cite{MEa05}. They have been extensively used to test theoretical predictions of MHD turbulence  (see, \cite{MT90,BB91,GVM91,PEa09,PB10,WEa11} and references therein).
The Elsasser fields, ${\bf Z}^\pm={\bf v}\pm{\bf B}/\sqrt{4\pi \rho}$, where ${\bf v}$ and ${\bf B}$ are the velocity and magnetic fields, respectively, and $\rho$ is the average density, represent eigenfunctions of counter propagating (with respect to the mean magnetic field) Alfv\'en waves and therefore they are primary fields for the study of incompressible MHD turbulence. Fluctuations in the fast solar wind are strongly imbalanced - there is more power in Alfv\'en waves propagating outward from the sun than toward it (e.g., \cite{BC05}) so that the power in ${\bf Z}^+$ dominates over that in ${\bf Z}^-$.
As with many other quantities that characterize physical properties of the turbulent flow (e.g. Yaglom relations \cite{PP98,SmEa09,CEa09} and dynamic alignment angle \cite{PEa09}), the Elsasser variables combine velocity and magnetic field fluctuations as a function of temporal scale. Pioneering observations from the HELIOS missions showed that in the fast solar wind streams at 1 AU the observed power spectrum of  ${\bf Z}^-$ (the subdominant component) did not follow a single power law shape. At very low frequencies ($f<3\times10^{-4}{\rm Hz}$) the spectral slope $\gamma_-$ was close to Kolomogorov's value ($\gamma_-\approx-1.67$), whereas at higher frequencies ($5\times10^{-4}{\rm Hz}<f<2\times10^{-3}{\rm Hz}$) the ${\bf Z}^-$ power spectrum was much shallower, with $\gamma_- \approx -1.3-1.4$ \cite{MT90,BB91,GVM91}. A similar trend was found more recently in WIND observations \cite{WEa11} where in the low frequency part of the inertial interval ($10^{-3}{\rm Hz}<f<10^{-2}{\rm Hz}$) ${\bf Z}^-$  nearly follows Kolmogorov scaling which at higher frequencies again is more shallow ($\gamma_-\approx-1.3$). The absence of single scaling of the subdominant Elsasser field in the inertial interval contradicts all recently developed models of strong, anisotropic imbalanced MHD turbulence \cite{LGS07,BL08,C08,PB09,PBh09} which predict a universal scaling for both dominant ${\bf Z}^+$ as well as  sub-dominant ${\bf Z}^-$ spectra. They also are inconsistent with the results of recent high resolution direct numerical simulations of imbalanced MHD turbulence which showed nearly the same spectral indices of the energy spectra in the inertial interval \cite{BL08,PB09}.

 Control of observational uncertainty in the in-situ observations is non-trivial, although these errors often have known bounds. There are different challenges for magnetic field and velocity measurements; solar wind velocity observations are intrinsically more uncertain compared to the magnetic field data \cite{McEa95}. In this Letter we propose a general, instrument independent method to estimate the uncertainty on velocity field fluctuations direct from the data. We obtain the systematic shift that this uncertainty introduces into observed spectral exponents.
 We will see that the shallower ${\bf Z}^-$ spectrum at high frequencies can be entirely accounted for by  this uncertainty in the velocity data and the observations of the ${\bf Z}^\pm$ spectra may in fact within achievable accuracy of the observations, be in agreement with the predictions of theory and numerical simulations.

We use data obtained by the WIND spacecraft at 3 second resolution.
Magnetic field data is provided by the MFI instrument \cite{LeEa95} and density and velocity data by the 3DP instrument \cite{LEa95}. We use observations  made during a quiet fast stream of April 04-06, 2008,  during which the solar wind speed remained above 550~km/s. The energy of compressive fluctuations was an order of magnitude lower than that of incompressible fluctuations and, consequently, magnetic and velocity fluctuations, being mainly Alfv\'enic, were dominated by the components perpendicular to the local mean field. The mean field, $\bar {\bf B}(t,\tau)$, at some time $t$ and on scale $\tau$ is defined as the magnetic field  averaged over the interval $[t-\tau, t+2\tau]$.  Similar to most studies of the  Alfv\'enic component of fluctuations (e.g., \cite{PEa09}) we focus on the perpendicular components of the fluctuations of the velocity defined as $\delta {\bf v}_\perp=\delta {\bf v} - (\bar {\bf B} \cdot \delta {\bf v})\delta {\bf v}$ and magnetic field $\delta {\bf B}_\perp=\delta {\bf B} - (\bar {\bf B} \cdot \delta {\bf B})\delta {\bf B}$ where $\delta {\bf v}(t,\tau)={\bf v}(t+\tau)-{\bf v}(t)$ and  $\delta {\bf B}(t,\tau)={\bf B}(t+\tau)-{\bf B}(t)$; in what follows
 subscript $\perp$ will be omitted for  simplicity.

In common with all velocity in situ observations, the 3 s velocity observations on WIND  (as well as any other in-situ measurements of moments of the particle distribution function) are quantized before ground transmission and this quantization results in high frequency noise or quantization noise \cite{McEa95}. This, and other contributions to observational uncertainty,  decorrelate the velocity and magnetic field fluctuations at high frequencies. White, delta correlated noise provides a reasonable generic, instrument independent model for the  uncertainty \cite{PB10}. Any measurement of a velocity component fluctuation $\delta v_o$ can then be represented as a sum of the 'real' turbulent signal $\delta v_s$ and a noise $\delta v_n$ which has zero mean and standard deviation $\varepsilon$, so $\delta v_o=\delta v_s + \delta v_n$. In what follows we will neglect the uncertainties in the magnetic field measurements since generally these  are small relative to that on the velocity measurements \cite{PB10,WEa11}.

 We will first quantify the velocity uncertainty from the data. We will exploit the fact that both the turbulent signal and the noise are random variables with distinct characteristic autocorrelation time scales. We make a key assumption- that the autocorrelation timescale of the underlying turbulent signal is that observed in the magnetic field component fluctuations $\delta B_o$ (they have negligible noise) and that this is also the autocorrelation timescale of the 'true' turbulent velocity component fluctuations $\delta v_s$. Any difference in the autocorrelation functions of the observed $\delta v_o$ and $\delta B_o$ are thus attributable to the (delta correlated) noise on the velocity $\delta v_n$.
 \begin{figure}
\begin{centering}
\includegraphics[width=1\columnwidth]{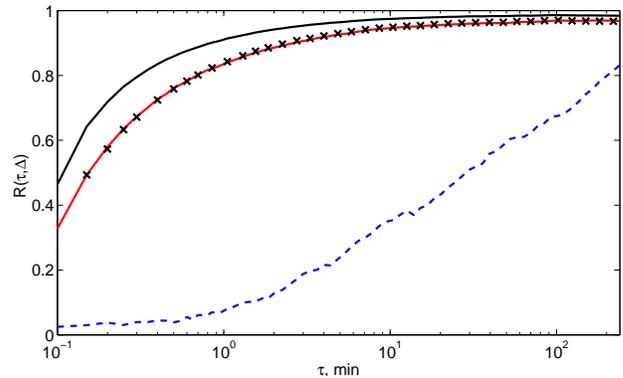}
\par\end{centering}
\caption{Autocorrelation functions $R_{\delta B_y}(\tau,\Delta \tau)$ (black line), $R_{\delta v_y}(\tau,\Delta \tau)$ (red line) and $R_{\delta v_p}(\tau,\Delta \tau)$ (dashed line), with the time lag $\Delta \tau = 3$ s.} \label{fig:fig1}
\end{figure}
 The autocorrelation coefficient (AC) of a component $\delta v$ on time lag $\Delta $ is  $R_{\delta v}(\tau, \Delta) \equiv E[(\delta v(t + \Delta,\tau)-\langle \delta  v \rangle) (\delta v(t, \tau)- \langle \delta  v \rangle)]/\sigma_{\delta v}^2$, where $E[~]$ is the expected value operator and $\sigma_{\delta v}$ is the standard deviation on a given velocity component.
The autocorrelation coefficients $R_{\delta B_o}(\tau, \Delta)$ and $R_{\delta v_o}(\tau, \Delta)$  are plotted for lag $\Delta=3{\rm s}$ as a function of scale $\tau$ with  black and red lines respectively in Figure 1. We see that the AC grows with scale $\tau$ for both signals and that the velocity AC is systematically lower that that of the magnetic field, consistent with the assumption of delta or uncorrelated noise  ($<\delta v_n(t+\Delta)\delta v_n(t)>=\delta(\Delta)$) that principally affects the velocity signal.
Given these assumptions one can construct a pseudo noisy signal by adding uncorrelated noise to the magnetic field observations. The pseudo noisy signal fluctuations $\delta B_{o+n}=\delta B_o+\delta B_n$, where $\delta B_n$ are delta correlated Gaussian distributed random numbers with zero mean and standard deviation $\varepsilon_B$. The magnitude of the pseudo noise  $\varepsilon_B$ can then be systematically varied and we plot on Figure 1 (black crosses) the result for a fractional uncertainty on the magnetic field corresponding to a velocity uncertainty of   $ \bar \varepsilon_B \equiv \varepsilon_B \sqrt{ \langle\delta v_o^2\rangle / \langle\delta B_{o+n}^2\rangle}=4~{\rm km/s}$. We see that this pseudo noisy signal closely coincides with the observed velocity AC, suggesting that $\varepsilon \sim 4~{\rm km/s}$ is a reasonable estimate of the amplitude of the noise on the turbulent velocity signal.  We will develop this idea to obtain  general methods to estimate the
 uncertainty direct from the data. First, we will see how these uncertainties can affect measurements of scaling exponents and the conclusions that can be drawn from them.
  \begin{figure}
\begin{centering}
\includegraphics[width=1\columnwidth]{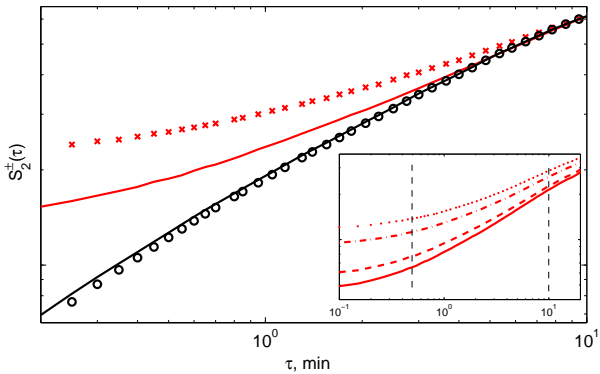}
\par\end{centering}
\caption{The normalized second order structure functions: of the $z$ component of the sub-dominant Elsasser variable $Z^-_z$ (red solid line), of $Z_z^-$ with added Gaussian noise with $\bar \varepsilon_B=4~{\rm km/s}$ (red crosses), of the dominant Elsasser variable $Z_z^+$ (blue solid  line) and $S_A$ (circles, see text for details). In the insert: the second order structure functions of the sub-dominant Elsasser variable for different values of added Gaussian noise. Raw observations are denoted by the solid line, and with added noise $\bar \varepsilon_B=2~{\rm km/s}$ (dashed line), $\bar \varepsilon_B=4~{\rm km/s}$ (dash-dotted line) and $\bar \varepsilon_B=5~{\rm km/s}$ (dotted line).} \label{fig:fig4}
\end{figure}

In Figure 2 we plot the observed second order structure functions $S_2$ of a component of the fluctuations in the Elsasser variables, where the Elsasser components are given by $\delta Z^\pm_i(\tau)=\delta v_i(t,\tau)\pm \delta B_i(t,\tau)/\sqrt{4\pi \bar \rho}$ and $\bar \rho(\tau)$ is the interval averaged local mean value of the density  over  time scale $\tau$, and $S_2^\pm =\langle \delta Z^\pm_i(\tau)^2\rangle$.
The solid lines are the structure functions of $z$ components of the dominant $\delta Z^+_z$ (black) and subdominant $\delta Z^-_z$ (red) fields.
They are normalized to have the same values at $\tau=10~{\rm min}$ scale on this plot; the power in $\delta Z^+_z$ is 20 times that in $\delta Z^-_z$.
 For the ideal statistical scaling of fully developed MHD turbulence we anticipate the scaling $S_2 \sim \tau^{\zeta_\pm(2)}$ and turbulence theories predict constant values of $\zeta^\pm(2)$ over the entire inertial interval (they are directly related to the power spectral exponents $\gamma_\pm$ via $\gamma_\pm=-\zeta_\pm(2)-1$).
 We can see that, consistent with earlier studies \cite{MT90,BB91,GVM91,WEa11}, the subdominant Elsasser variable does not follow a single power law in the inertial interval. A linear mean least square fit over scales $30~{\rm s}<\tau < 10~{\rm min}$ gives $\gamma_+=-1.54\pm 0.02$ and $\gamma_-=-1.40\pm0.02$, consistent with previous observations.
 
 We now test the idea that the uncertainty in the observed velocity, estimated above, is sufficient to account for this observed departure from the theoretical predictions. Since the power in  $\delta {\bf Z}^+$  is significantly higher than that in $\delta {\bf Z}^-$ we will focus on the effects of uncertainties in velocity on the $\delta {\bf Z}^-$ signal only.
 We calculate $S_2^-(\delta B_{o+n}, \delta v_o)$ using the pseudo noisy signal fluctuations $\delta B_{o+n}$ and the observed $\delta v_o$. This is shown in the inset of Figure 2 for a range of values of $\bar \varepsilon_B$. We can see that addition of  'white' (delta correlated) noise always systematically 'flattens'  these curves, that is, it decreases the value of the scaling exponent; for $\bar \varepsilon_B=4~{\rm km/s}$  pseudo noise strongly affects $S_2^-$ at all scales in the inertial interval. The 'flattening' of the pseudo-noisy $S_2^-(\delta B_{o+n}, \delta v_o)$ curve, that is, the change in  the mean exponent over timescales $30~{\rm s}<\tau < 10~{\rm min}$
is $\Delta \gamma_-\approx 0.13$,  is close to  the observed difference between exponents of the dominant and subdoninant fields ($\gamma_- - \gamma_+=0.14$) hence this difference could be just due to noise in the velocity data.
 
The pseudo noisy  $S_2^-$ curve generated
 with noise $\bar \varepsilon_B=4~{\rm km/s}$  is plotted as the red crosses in the main panel of the Figure. Since the noise is assumed to be linearly additive, the difference between the observed, and the pseudo noisy $S_2^-$ curves, that is, $\epsilon_S=[S_2^-(\delta B_{o+n}, \delta v_o)- S_2^-(\delta B_o, \delta v_o)]$ provides an estimate of how a velocity uncertainty of $ \varepsilon=4~{\rm km/s}$ affects the subdominant Elsasser variable scale by scale. We then compensate for this systematic effect by  subtracting this $\tau$ dependent uncertainty from the observed $S_2^-(\delta B_o, \delta v_o)$ and $S_2^-(\delta B_o, \delta v_o)-\epsilon_S$ is shown by the black circles on the plot.
  This compensated subdominant $S_2^-$ curve now has a single scaling range, consistent with current theories and numerical predictions \cite{LGS07,BL08,C08,PB09,PBh09}. It can also be seen to be in remarkable agreement with that observed for the dominant Elsasser variable. The uncertainty in the velocity that we have estimated from the data, as shown in Figure 1, is thus sufficient to account for the departure in scaling between the $\delta{\bf Z}^-$ and $\delta{\bf Z}^+$ Elsasser variables and these observations may in fact within the achievable accuracy  be in agreement with theories \cite{LGS07,PB09,PBh09} that predict a \emph{single scaling} for $\delta{\bf Z}^-$ and $\delta{\bf Z}^+$.

We have verified that $\delta {\bf Z}^+$ is not strongly affected by velocity uncertainties of this amplitude:  $S_2^+$ remains almost unchanged for $\bar \varepsilon_B<5{\rm km/s}$. This conclusion is also supported by the fact that as shown in Figure 2, $S_2^+$ has a convex shape, which is typical for finite range hydrodynamic and MHD turbulence \cite{BEa93,CN09},  whereas $S_2^-$  is concave, consistent with 'shallowing' at small scales due to noise effects.
In addition to the velocity measurement uncertainties the Elsasser fields are also affected by uncertainties related to the density measurement. Our analysis shows that adding the same amount of relative pseudo noise to the density data had negligible influence on the results. This is not surprising, as the definition of the Elsasser fields is in terms at the mean value of the density over scales, and this reduces the influence of density uncertainties due to the central limit theorem.

We now outline methods to
 explicitly determine the uncertainty as a function of scale $\tau$ from the observations. We will generalize the approach shown in Figure 1.
Given the assumption that $R_{\delta B_o}(\tau,\Delta)=R_{\delta v_s}(\tau,\Delta)$ and again that the noise is delta correlated $<\delta v_n(t+\Delta)\delta v_n(t)>=0$ we can obtain the uncertainty from the AC directly:
\begin{equation}
\varepsilon = \sigma_{\delta v_o} \sqrt{1-\frac{R_{\delta v_o}(\tau, \Delta )}{R_{\delta B_o}(\tau,\Delta)}}.\label{eq:AC1}
\end{equation}
% where $R_{\delta B_o}(\tau,\Delta)>R_{\delta v_o}(\tau,\Delta)$ as above.
Alternatively, we can
 estimate the uncertainty in velocity fluctuations using a pseudo noisy signal constructed by adding uncorrelated noise to the magnetic field observations: $\delta B_{o+n}=\delta B_o+\delta B_n$. We can vary $\varepsilon_B$ until the AC of the pseudo noisy magnetic field $R_{\delta B_{o+n}}(\tau,\Delta)$  coincides with that of the velocity $R_{\delta v_o}(\tau,\Delta)$. The relation $\varepsilon_B(\Delta)/\sqrt{\langle\delta B_{o+n}^2\rangle} = \varepsilon(\Delta )/\sqrt{\langle\delta v_o^2\rangle}$  provides a  scale dependent estimate of $\varepsilon$.
These methods are demonstrated in Figure 3 and give results that are consistent with the estimate of $\varepsilon=4~{\rm km/s}$ used above.

Recently, in-situ solar wind data have been used \cite{PEa09} to test the prediction of scale dependent dynamic alignment
in MHD turbulence \cite{B06}. This relies on determination of the angle between magnetic and velocity fluctuations perpendicular to the local mean magnetic field direction. The observational uncertainty is known to make a significant contribution to the component of the velocity perturbation perpendicular to the magnetic field perturbation $\delta {\bf v}_p(t,\tau) = \delta {\bf B}_\perp \times \delta {\bf v}_\perp/\delta B_\perp$  even at quite large scales $\tau \sim 10~{\rm min}$ \cite{PEa09}. We show this in Figure 1, where the AC of $\delta  v_p(t,\tau)$ is plotted (dashed line), we can see that the magnitude of the AC is much lower than that of the components of velocity and magnetic field. This AC function allows us to develop one more method to determine measurement uncertainty. Representing $\delta v_{p,o}$ as a sum of the underlying turbulent signal and a noise $\delta v_{p,o}=\delta v_{p,s}+\delta v_{p,n}$ and assuming that the AC of $\delta v_{p,s}$ and the magnetic field fluctuations are identical, we have that:
\begin{equation}
\varepsilon = \sigma_{\delta v_{p,o}} \sqrt{1-\frac{R_{\delta v_{p,o}}(\tau, \Delta )}{R_{\delta B_o}(\tau,\Delta)}}.\label{eq:AC2}
\end{equation}
The scale dependent estimate of $\varepsilon$ derived using this approach is given by the red dash-dotted line in Figure 3 and the result is in very good agreement with the results obtained by the other methods. We also plot the rms value of the observed $\delta v_{p,o}$ on Figure 3 (black line) and this can be seen to coincide with our various estimates of the \emph{uncertainty} in the velocity fluctuations on scales $\tau$ of a few minutes or less. On these smallest scales, the observed $\delta {\bf v}_{p,o}$ is almost entirely dominated by observational uncertainty.
 
 The  estimate of the quantization error  in  \cite{PEa09}  gave a somewhat lower value $\varepsilon \approx 1.5~{\rm km/s}$. In \cite{PEa09}  the alignment angle  $\theta_A \equiv \arcsin (\langle \delta v_p  \delta B_\perp  \rangle/\langle \delta v_\perp \delta B_\perp  \rangle) \approx v_p/\sqrt{\langle \delta v_\perp(\tau)^2\rangle}$ was used to estimate the quantization error. In fact correlation between $\delta B_\perp$ and $\delta v_\perp$ is much stronger than correlation between $\delta B_\perp$ and $\delta v_p$ (because the latter at small scales is strongly dominated by the error). Detailed analysis shows that this increases  the estimates in \cite{PEa09} by a factor $\sim 1.6$. Also, the local alignment angle between velocity and magnetic fields $\theta$ is random variable with zero mean. Any measure of the alignment angle (such as mean of the absolute value of the alignment angle) should thus be related to the standard deviation $\sigma_\theta$ of the angle. If for highly aligned cases we assume that $\theta$ is close to Gaussian distributed, then its absolute value is described by the half-normal distribution and its mathematical expectation is given by $\sigma_\theta \sqrt{1-2/\pi}$ (and not $\sigma_\theta$) and this yields another factor of $1.2533$ compared to the estimate of \cite{PEa09}.

This uncertainty of $\approx 4~{\rm km/s}$ in the velocity fluctuations will systematically reduce the scaling exponent of the velocity compared to that of the magnetic field, if as we have assumed here, the uncertainty in the magnetic field can be neglected in such a comparison. We estimate that this uncertainty alone would result in a difference in the power spectral exponents of $\gamma_v - \gamma_B \approx 0.04$. This is less than the observed difference which is typically in the range $\sim 0.1$ \cite{PB09} to $\sim 0.2$ \cite{TEa09}. These estimates are sufficiently close however to merit more detailed analysis.

In summary, we have presented general, instrument independent methods to determine the uncertainty in velocity fluctuations in single point measurements. We have shown that this uncertainty is sufficient to account for the departure in scaling between the subdominant and dominant Elsasser variables and thus are able to report for the first time that the observations
  are, within the achievable accuracy, in agreement with the predictions of theory and numerical simulations. Our results put careful estimation of uncertainties at the centre of the testability of theoretical predictions of scaling exponents.  Our approach, and development of it, is highly relevant for methods developed
for the study of MHD energy cascade rates in the solar wind. These inevitably involve combinations of velocity and magnetic field fluctuations that are scale dependent such as mixed third order moments of the Elsasser fields \cite{SmEa09,CEa09} via Yaglom relations \cite{PP98}.
\begin{figure}
\begin{centering}
\includegraphics[width=1\columnwidth]{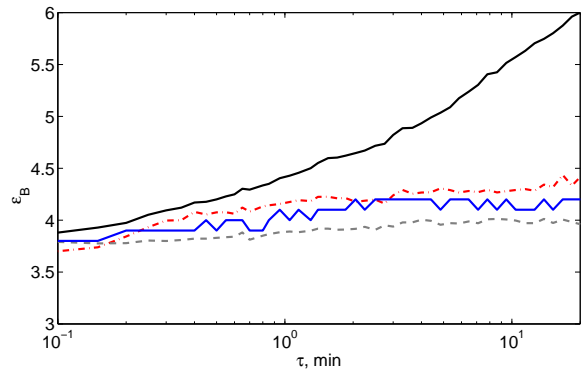}
\par\end{centering}
\caption{Quantization error derived by different methods (see text for details): green line correspond to the result derived usimg Eq. (\ref{eq:AC1}), blue line is derived by adding artificial noise to the magnetic field data. Red line corresponds to the result derived using Eq. (\ref{eq:AC2}). Black line denotes rms value of $\delta v_p$.} \label{fig:figN}
\end{figure}
\begin{acknowledgments}
The authors acknowledge the WIND instrument teams for providing MFI and 3DP data. This work was supported by the UK STFC.
\end{acknowledgments}

\end{document}